\documentclass[aip, showpacs]{revtex4}

\usepackage{amsmath}
\usepackage{amssymb}
\usepackage{graphicx}
\usepackage{bm}
\usepackage{epstopdf}
\usepackage{graphicx,epstopdf}
\usepackage{color}

\newcommand{\la}{\lambda}

\def\be{\begin{equation}}
\def\ee{\end{equation}}
\begin{document}

\title{Critical density of a soliton gas}

\author{G. A. El}
\email{g.el@lboro.ac.uk}
\affiliation{Department of Mathematical Sciences, Loughborough
University, Loughborough LE11 3TU, UK}

\begin{abstract}
We quantify the notion of a dense soliton gas by  establishing an upper bound for the integrated density of states of the  quantum-mechanical Schr\"odinger operator  associated with the KdV soliton gas dynamics.  As a by-product of our derivation we find  the speed of sound in the soliton gas with Gaussian spectral distribution function.
\end{abstract}

\pacs{05.45.Yv, 02.30.Ik, 92.10.Hm}

\maketitle

\section{Introduction}

Dynamics of incoherent nonlinear dispersive waves have  recently become the subject of a very active research in nonlinear physics, most notably in  oceanography, nonlinear optics and condensed matter physics.
In some cases such dynamics can be viewed as a natural  counterpart of turbulent motion in traditional dissipative fluid systems (see e.g. \cite{turitsyn_nature}). In the context of dispersive wave motion, turbulence is usually associated with a complex, spatio-temporal  wave dynamics that requires a statistical description, the prominent example being the wave turbulence  theory pioneered by V.E. Zakharov \cite{zakharov67}. The extension of the notion of turbulence to dispersive wave systems is particularly compelling when the governing system of equations is integrable, which provides one with the principal availability of the full analytical description \cite{lax91}. The emerging theory of  integrable turbulence  \cite{zakharov09}  encompasses both weak (wave) and strong (soliton) turbulence. The  description of integrable wave turbulence  has found recent development in \cite{pierre, stephane1} where  the theoretical findings were confirmed  in the fibre optics experiments. The `opposite' case of integrable soliton turbulence has a much longer history dating back to 1971 Zakharov's paper on the kinetic equation for solitons \cite{zakharov71} but the experimental observation of  soliton gas/soliton turbulence in shallow water ocean waves has been reported only very recently \cite{osborne2014} and revealed striking low frequency power law Fourier spectra of the measured random nonlinear wave field. 

To be clear from the very beginning,  the notions of soliton gas and (integrable) soliton turbulence, at least the way they are used in this paper, represent two complementary aspects of the same physical object. These two aspects are the natural counterparts of the particle-wave duality of a single soliton. In the soliton-gas description the focus is on the collective dynamics of solitons as interacting particles  characterised by a certain amplitude (velocity) distribution function, while the soliton turbulence description emphasises the properties of the random nonlinear wave field associated with the soliton gas.  In this paper we consider certain properties of the soliton gas/soliton turbulence for the Korteweg -- de Vries (KdV) equation
\be\label{kdv}
u_t-6uu_x+u_{xxx}=0 \, .
\ee
Despite the  deceptively `old-fasioned' nature of this object there still are a number of open fundamental questions pertaining to the behaviour of random solutions to equation (\ref{kdv}) (see e.g. \cite{kotani}, \cite{zakharov2015}). 

The inverse scattering theory associates each KdV soliton  with a point of discrete spectrum $\la_n=-\eta^2_n$ of the Schr\"odinger operator
\be\label{schr}
\mathcal{L}=-\partial_{xx}^2  + u(x,t)\, .
\ee
Along with the spectral parameter $\eta_j>0$, each soliton is characterised by  the `phase' $x_j \in (-\infty, + \infty)$ determining its spatial location (not necessarily coinciding with the position of the local maximum of $u(x,t)$). In a soliton gas, the spectral parameter $\eta_n$  is distributed on a finite interval $\mathfrak{S} \subset {\mathbb R}$ (which without loss of generality can be assumed to be $[0,1]$) with some density $\phi(\eta)$, while the individual soliton locations $x_j$ have the Poisson distribution on the line with some density  parameter $\kappa$ \cite{ekmv2001}. Thus, mathematically, soliton gas can be viewed as a compound Poisson process \cite{feller}.
The spectral distribution function  $f(\eta)$  of solitons in the gas  is then introduced  such that $f (\eta_0) d\eta dx$ is the number of solitons with the spectral parameter $\eta_n \in (\eta_0 , \eta_0 + d\eta)$ found in the space interval $(x, x+dx)$ at the moment of time $t$, i.e. $f(\eta)$ is the density of states per unit length.  The integral density $\kappa$ of the soliton gas is found as 
\begin{equation}\label{kappa}
\kappa = \int_{\mathfrak{\mathfrak{S}}} f(\eta) d \eta \, .
\end{equation}
In an inhomogeneous soliton gas one has $f=f(\eta; x, t)$. The evolution of $f$ is then governed by the kinetic equation first derived by Zakharov for the case of a `rarefied'   ($\kappa  \ll 1$) soliton gas \cite{zakharov71}. Zakharov's kinetic  equation was generalised in  \cite{el03}, \cite{ek05}  to the case of soliton gas of arbitrary ($\kappa= O(1)$) density.
This non-perturbative kinetic equation for a `dense' soliton gas has the form 
\begin{eqnarray} 
&& f_t+(sf)_x = 0\, ,  \label{kin1} \\
&&s(\eta)=4\eta^2+\frac{1}{\eta}\int _{\mathfrak{S}} \log
\left|\frac{\eta + \mu}{\eta-\mu}\right|f(\mu)[s(\eta)-s(\mu)]d\mu\, .  \nonumber
\end{eqnarray}
Here  we used the shorthand notation $f(\eta) \equiv f(\eta; x, t)$,  $s(\eta) \equiv s(\eta; x, t)$, the latter being  the mean, or effective,  velocity of a soliton with the spectral parameter $\eta$ in a soliton gas, which differs, owing to soliton interactions, from the free soliton velocity $4 \eta^2$. 
The typical scales of $x$ and $t$ in (\ref{kin1}) are much larger than in the KdV equation (\ref{kdv}).

Equation (\ref{kin1}) provides kinetic description of a dense (as opposed to rarefied) soliton gas in the sense that the second term in the integral equation in (\ref{kin1}) describing soliton interactions has generally the same  order as the first term related to the free soliton motion. At the same time, this equation   does not  impose any specific limitations  on the density (except for its boundedness) and does not imply any qualitative changes in the KdV solution behaviour due to large density values.    It appears, however,  from the observational results of \cite{osborne2014} that the density parameter plays crucial role in the formation of the power-law Fourier spectra of the KdV soliton turbulence. Indeed, the Fourier spectra of the shallow water soliton turbulence observed in \cite{osborne2014} exhibit the power-law behaviour $\omega^{-1}$, while the spectra of the rarefied soliton gas are  exponential \cite{gurzybel99}. The energy shift to lower frequencies clearly occurs due to soliton interactions whose role increases with the increase of the gas density.  Thus, an additional consideration is required in order to understand the effect of the soliton gas density on the properties of the associated soliton turbulence.

The main result of this  paper is the establishment of an upper bound for the density of a KdV soliton gas, so that the notion of dense soliton gas acquires the definitive quantitative criterion. 
As a by-product of our calculation, we  show that the evolution of the density disturbances in a soliton gas satisfies the linear transport equation  whose characteristic velocity yields  the speed of sound in  a soliton gas.

\section{Spectral distribution function and the  moments of soliton gas}

\subsection{Finite-gap potentials: quasi-momentum and the integrated density of states}

The spectral distribution function $f(\eta)$ for  the KdV soliton gas is most naturally defined  in terms of the integrated density of states  of the associated Schr\"odinger operator (\ref{schr}) 
\cite{lgp}:
\begin{equation}\label{ds}
{\cal N}(\lambda)=\lim \limits_{b-a \to \infty}\frac{\nu(a,b;\lambda)}{b-a} \, ,
\end{equation}
where $\nu(a,b;\la)$ is the number of eigenvalues $\la_j \le \la$ in the Dirichlet problem $(-\partial_{xx}^2+u(x,t))\psi = \lambda \psi$ on $a\le x \le b$: $\psi(a, \lambda)=\psi(b, \lambda)=0$. 
It is known \cite{JM} that for almost-periodic potentials $u(x)$ the differential $d{\cal N}$ is a measure supported on the spectrum so one can introduce the distribution function $f(\eta) >0$ such that  $d {\cal N}=f(\eta) d \eta$, where $\eta^2 = -\lambda$.  We shall be interested in the properties of  this spectral measure  for soliton gas, which can be   constructed as the {\it thermodynamic limit} of finite-gap potentials \cite{ekmv2001}.

The finite-gap potentials   play  the prominent role in the KdV theory (see \cite{TS} and references therein). Remarkably, the  corresponding Schr\"odinger operators  have the spectrum consisting of $N \in \mathbb{N}$  disjoint finite bands $[\lambda_{2i-1}, \lambda_{2i}]$ plus one semi-infinite band $[\lambda_{2N+1}, \infty)$. The bands are seprarated by $N$ finite gaps. 
Shrinking a finite band into a point corresponds to the appearance of a soliton on the ($N-1$)-gap potential `background' \cite{TS}. Collapsing all $N$ bands into points yields the $N$-soliton KdV solution.  

For a $N$-gap potential  the integrated density of states (\ref{ds}) can be calculated as ${\cal N}_N(\lambda)= \frac{1}{\pi} Re [p_N(\la)]$, where $p_N(\lambda)$  is  the {\it quasi-momentum}, a fundamental quantity with the well-defined analytic properties \cite{TS, dn89}.  For our consideration it is sufficient to know that the quasimomentum is the generating function for the averaged {\it Kruskal integrals},  the densities of the KdV conservation laws averaged over the family of $N$-gap solutions, so that the expansion  of ${p}_N(\lambda)$ near $\lambda = \infty$ has the form \cite{TS, dn89}
\begin{equation} \label{10} 
p_N = \sqrt{\lambda}+\sum
\limits^{\infty}_{k=0}\frac{I_k^{(N)}}{(2\sqrt{\lambda})^{2k+1}} \, , \qquad \lambda \gg 1,
\end{equation}
where, in particular, 
\begin{equation} \label{kr}
 I_0^{(N)}=\lim \limits_{L \to
\infty} \frac{1}{L} \int \limits ^L_0 u_N(x)dx \, , \quad I_1^{(N)}=\lim \limits_{L \to \infty} \frac{1}{L}\int \limits ^L_0
u_N^2(x)dx \, .
\end{equation}
(We note that  $N$-gap potentials $u_N$ are quasiperiodic functions, so the existence of the averages in (\ref{kr}) is guaranteed). The higher order averaged Kruskal integrals don't coincide with the higher moments of the wave field so they won't be used in what follows.

We  shall use  expressions (\ref{kr}), (\ref{10})  to compute the moments $I_0= \overline u  $, $I_1= \overline {u^2}$ of the soliton gas in terms of the spectral distribution function $f(\eta)$ introduced formally in the previous section. For that, we need to briefly outline the KdV soliton gas construction proposed  in \cite{ekmv2001}, \cite{el03} which  provides the connection between the  quasi-momentum differential  $d p_N(\lambda)$ in the limit as $N \to \infty$ and the spectral distribution $f(\eta)$.

\subsection{Soliton gas  construction}
Consider  a sequence of $N$-gap solutions   $u_N(x,t)$, $N=1, 2, 3, \dots $ 
of the KdV equation (\ref{kdv}), defined in the following way. Let the finite-band spectrum of $u_N$  be confined to some given interval, say $ [-1, 0]$ (without loss of generality we can set $\la_{2N+1}=0$). The $N$-gap potentials are multiphase ($N$-phase) KdV solutions so that $u_N(x,t)=U_N(\theta_1, \theta_2, \dots, \theta_N)$, where $\theta_j=k_jx + \omega_j t + \theta_j^{(0)}$, $k_j$ being the wavenumbers, $\omega_j$ the frequencies, and $\theta_j^{(0)}$ the initial phases. 
The quantities $k_j$ and $\omega_j$ are defined in terms of the spectrum edges $\{ \la_j \}_{j=1}^{2N+1}$ (see e.g. \cite{dn89}) and $\theta_j^{(0)}$ are arbitrary.  In particular,
\be \label{kj}
k_j= 2 \int \limits_{\la_{2j-1}}^{\la_{2j}} dp_N(\la), \quad j=1, \dots, N.
\ee
The total integrated density of states in $N$-gap potential can then be found as (see \cite{JM})
\be \label{total}
{\cal N}_N(0) = \frac{1}{\pi} Re  \int_{-1}^0 dp_N (\la)= \frac{1}{2\pi} \sum_{j=1}^N  k_j ,
\ee
i.e. it has the physically transparent meaning of the total `density of waves'.

The functions $U_N(\theta_1, \dots, \theta_N)$ are $2\pi$-periodic  with respect to each phase $\theta_j$ and therefore, $u_N(x,t)$ are quasi-periodic 
in both $x$ and $t$ provided the wavenumbers $k_j$ as well as frequences $\omega_j$ are incommensurate.
The soliton gas construction  then  proceeds as follows \cite{ekmv2001}:  (i) assume that the initial phases $\theta_j^{(0)}$ are independent random values uniformly distributed on $[-\pi , \pi)$, i.e. the vector $\boldsymbol{\theta}^{(0)}$ is uniformly distributed on the $N$-torus \cite{osb93}; (ii) consider the  sequence of finite gap potentials $u_N(x,t)$ such that $N \to \infty$ but  the total integrated density of states is fixed, i.e. ${\cal N}_N (0) = O(1)$.   The corresponding limit of finite-gap potentials (assuming its existence in some (weak) sense) represents thus  an analog of the thermodynamic limit in statistical mechanics.

The requirement of boundedness of the total density of states (\ref{total}) in the thermodynamic limit as $N \to \infty$ implies certain asymptotic structure (scaling) of the finite-band spectrum.   Indeed, the existence of  $\lim \limits_{N \to \infty}{\cal N}_N(0)< \infty$  implies  $k_j =O(N^{-1})$ for $N \gg 1$. The analysis of (\ref{kj}) then  yields that the spectral bands and gaps of $u_N(x)$ for $N \gg 1$   are distributed in such a way that 
\be \label{scaling1}
|{\hbox{gap}_j}| =  \la_{2j+1} - \la_{2j}  \sim  \frac{1} {\phi(\eta_j) N};   \quad   |\hbox{band}_j|  = \la_{2j} - \la_{2j-1} \sim \exp{(- \gamma(\eta_j) N)}, 
 \ee
where $\eta_j=\frac{\la_{2j-1} + \la_{2j}}{2}$ is the centre of the $j$-th band,  and $\phi(\eta)$, $\gamma(\eta)$ are some  continuous functions on $[0,1]$ (see \cite{ven90}, \cite{el03} for details). Then we have for the limit of the band-gap ratio: 
\be \label{tscaling}
\frac{|\hbox{band}_j|}{|{\hbox{gap}_j}|} \sim N \exp{(- \gamma N)} \to 0  \quad \hbox{as} \ \ N \to \infty,
\ee
for all $j$, which corresponds to the soliton (more precisely, infinite-soliton) limit.  

Having defined the thermodynamic limit for the spectrum of finite-gap potentials, we need now to determine what happens in this limit with the random phases $\theta_j=k_jx + \omega_jt+\theta^{(0)}_j$, $j=1, \dots, N$. The scaling (\ref{scaling1}) implies that in the thermodynamic (infinite-soliton)  limit all the wavenumbers and frequencies vanish,  $k_j \to 0$, $\omega_j \to 0$, i.e. the spatial and temporal periods become infinite. The latter implies that the  phase torus in the thermodynamic limit  maps onto  the infinite line \cite{ekmv2001}.  To show that, one  represents the phases $\theta_j \in [-\pi, \pi)$ in the form $\theta_j=k_j(x - x_j)$, where  $x_j=\theta_j^{(0)}/k_j \in [-\pi/k_j,  \pi/k_j)$ (the temporal components $\omega_j t$ of the respective phases are assumed to be absorbed in the random initial phases $\theta_j^{(0)}$).  Then it is not difficult to show \cite{ekmv2001} that the uniform distribution of $\boldsymbol {\theta}^{(0)}$ on $[-\pi, \pi)^N$ transforms under  the thermodynamic limit  into the Poisson distribution for the ``soliton centres'' $x_j$ on $(-\infty, \infty)$. 
The  dynamics of the spatial phases  $\xi_j=x-4\eta_j^2 t -x_j$, $j \in \mathbb{Z}$ in soliton gas  are thus equivalent to the particle dynamics in an ideal configuration gas constructed as the thermodynamic limit of the dynamical system of $N$ particles moving with constant speeds on a circle, see e.g. \cite{sinai}.

\subsection{Moments of soliton turbulence}
We now  express the thermodynamic limit of the moments (\ref{kr}) of the nonlinear wave field $u(x,t)$ in the soliton gas  (i.e. soliton turbulence)  in terms of the averages over the limiting  spectral measure $d {\cal N}_\infty \equiv f(\eta) d\eta$, where $f(\eta)$ is the spectral distribution function of the soiton gas. For that we first use the $\lambda$-derivative of the expansion (\ref{10}) to obtain:
\be \label{mom0}
I_0^{(N)} = -4 \hbox{Res}_{\la=\infty} [ \la^{1/2}\frac{dp_N}{d \lambda}]= -\frac{2}{ \pi i} \oint \limits_{C_{\infty}} \la^{1/2} dp_N .
\ee
Here $C_{\infty}$ is the contour  surrounding the point $\la=\infty$ clockwise.
Similarly,
\be \label{mom1}
I_1^{(N)} = -\frac{16}{3} \hbox{Res}_{\la=\infty} [ \la^{3/2}\frac{dp_N}{d \lambda} ]= - \frac{8}{3\pi i} \oint \limits_{C_{\infty}}\la^{3/2}  dp_N(\la).
\ee
 It is  not difficult to show using the properties of the quasi-momentum (see e.g. \cite{dn89})   that for the thermodynamic scaling (\ref{tscaling}) $Im (p_N) \to 0$ as $N \to \infty$, and so the spectral distribution function $f(\eta)$ of soliton turbulence is found as \cite{el03, el_jns}
\be\label{sd}
 f(\eta) d \eta \equiv \lim_{N \to \infty}  d {\cal N}_N =  \frac{1}{\pi}\lim_{N \to \infty} d p_N.
\ee
Now, applying the thermodynamic limit to (\ref{mom0}), (\ref{mom1}) and making the integration contour deformation $\oint \limits_{C_{\infty}}...d \la = \oint \limits_{\alpha}...d \la $, where $\alpha$ is the contour surrounding the spectral interval  $\la =-\eta^2 \in [-1, 0]$ counterclockwise, we obtain on using (\ref{sd}), the expressions for the two first moments in the KdV soliton turbulence 
\begin{equation}\label{mean}
\begin{split}
\overline{u}= - 4\int \limits_{0}^{1}\eta f(\eta )d\eta  = -4 \kappa \langle \eta \rangle \,, \quad
\overline{u^{2}}
=\dfrac{16}{3}\int \limits_{0}^{1 }\eta ^{3}f(\eta)d\eta  = \frac{16}{3} \kappa \langle \eta^3 \rangle \, ,
\end{split}
\end{equation}
where the averaging over space is defined by (\ref{kr}) and the angular brackets denote the averaging over the spectral distribution function  $f(\eta)$,
\begin{equation}\label{smom} 
\langle \eta^n \rangle = \frac{1}{\kappa} \int_{0}^{1 }\eta^n
f(\eta)d\eta \, , \quad n=1,2, \dots, \quad \kappa = \int \limits_0^1 f(\eta) d\eta.
\end{equation}

We  note that expressions (\ref{mean})  coincide with the expressions for the moments computed for  a soliton lattice , $-\sum  2\eta_j^2 \hbox{sech}^2[\eta_j(x-4\eta_j^2t -x_j)]$, see Ref. \cite{dutykh}. The results of Ref. \cite{dutykh} correspond to the rarefied gas limit  $\kappa \ll 1$ while for the dense soliton gas case studied here one generally has $\kappa=O(1)$. The mentioned coincidence is, however, not that surprising  as the density parameter $\kappa$ enters the full expressions  for the moments  (\ref{mean})  as a factor  so their form is retained in the asymptotic limit  $\kappa \ll 1$.

\section{Critical density of a soliton gas}
We consider the variance function of the KdV soliton turbulence,
\begin{equation}\label{var}
\mathcal{A}^2= \overline {u^2} - \overline u ^2 \ge 0,
\end{equation}
which is a measure of the integral intensity of fluctuations of the nonlinear turbulent wave field relative to its
mean $\overline u$. We now use the connection (\ref{mean}) between the spatial and spectral moments to see 
the possible restrictions  imposed on the spectral distribution function $f(\eta)$  by non-negativity of $\mathcal{A}^2$.

We first  consider the simplest, one-component `cold' soliton gas
characterised by the delta-function distribution function 
\begin{equation}\label{an1}
f(\eta) = f_0 \delta (\eta - \eta_0)\, ,
\end{equation}
where $\eta_0$ is the spectral parameter of the component and  the soliton gas density (\ref{kappa}) $\kappa=f_0$.
We substitute the ansatz (\ref{an1}) into the
expressions for the moments (\ref{mean}) to obtain
\begin{equation}\label{mean1}
\overline u = - 4\eta_0 \kappa\, , \qquad
\overline{u^2}=\frac{16}{3}\eta_0^3 \kappa\, ,
\end{equation}
which yields the variance function (\ref{var})
\begin{equation}\label{var1}
\mathcal{A}^2 = 16 \kappa \eta_0^2(\frac{\eta_0}{3} - \kappa).
\end{equation}
Now one can see that non-negativity of the variance (\ref{var1}) imposes a restriction on the
possible  values of the soliton gas density: 
\begin{equation}\label{crit}
\kappa  \le  \kappa_{cr}=\frac{\eta_0}{3}.
\end{equation}
At  $\kappa  = \kappa_{cr}$ one has  $\mathcal{A}^2=0$,  hence $\overline {u^2} = \overline u ^2$  which implies the absence of small-scale fluctuations.  
On the other hand,  it follows from (\ref{var1}) that for a given $\eta_0$ the maximum of the intensity of
fluctuations is achieved when the gas  density $\kappa = \eta_0/6$.  

We note that the same expression (\ref{crit}) for the  critical density of the cold gas was obtained in \cite{shurgalina} 
by  the formal computation of the condition $\mathcal{A}^2=0$ for a lattice of non-interacting solitons. Although this model is not applicable to the description of a dense gas, where interactions between soltons are essential,  it yields the same formula due to  already mentioned factorized structure of the full expressions for the moments (\ref{mean}).   

One can trace an instructive analogy between the critical parameter
(\ref{crit}) in a one-component soliton gas and the maximum of the density  of waves $k_0/(2\pi)$ in the 
 KdV dispersive shock wave (DSW), where $k_0$ is the wavenumber at the DSW trailing edge, where the amplitude of the
small-scale oscillations vanishes and one has $\overline{u^2} =\overline{u}^2$ (see \cite{GP},\cite{el05}). The analogy  is supported by  the well known fact
  \cite{lax_lev, ven90, llv94}  that the process of the generation of a
DSW can be described in terms of the asymptotic evolution of
a nearly reflectionless potential approximated by a $N$-soliton solution of the KdV
equation with $N \gg 1$. 
Assuming the initial condition for the KdV equation (\ref{kdv}) in the form of a wide rectangular well of a width $L \gg 1$ and depth $\Delta = O(1)$ 
the resulting DSW can be represented as the result of a coherent interaction of a large number of solitons having nearly the same spectral parameter 
$\eta_0=\sqrt{\Delta}$, and so can be viewed as  a coherent counterpart of the single-component soliton gas with the density gradually decreasing from the value $\kappa_0= k_0/(2\pi)$
at the trailing edge to $\kappa =0$ at the leading edge. It is known that
$k_0 = 2\sqrt{\Delta}$, where $\Delta$  is the jump across the DSW (see \cite{GP}, \cite{el05} with the account of a different normalisation of the KdV equation compared to (\ref{kdv})). The wave density at the harmonic edge is then
 $\kappa_0= \eta_0/\pi$ and has a natural interpretation  as the maximum of the density of solitons.  
It is interesting to note that the obtained maximum density of solitons in a DSW  is just below  the value $\kappa_{cr}= \eta_0/3$ (\ref{crit}) in the
counterpart cold  soliton gas with the distribution function (\ref{an1}).
 
We now consider  the soliton gas with the Gaussian spectral distribution function 
\begin{equation} \label{gauss}
f(\eta)=\frac{f_0}{\sqrt{2 \pi \sigma^2}} \exp\{-\frac{(\eta - \eta_0)^2}{2\sigma^2}\},
\end{equation}
where $\sigma^2$ is the spectral variance. Since $f(\eta)$ (\ref{gauss}) is defined for all $\eta \ge 0$ one can without loss of generality use $+\infty$ as the upper limit in all integrals over the spectrum $\mathfrak{S}$. Assume $ \eta_0=O(1)$ and $ \sigma \ll \eta_0$ so that  the contribution of the non-physical, negative   values of $\eta$ can be neglected and the normalisation $ \kappa = \int_0^\infty f(\eta)d\eta = f_0 $  remains (approximately)  valid.  
 The value of $\sigma^2$ can be interpreted as the measure of the soliton gas `temperature', characterising the spread of the  spectral parameter $\eta$ (and hence, soliton velocity)  around the dominant value $\eta_0$ ($4\eta_0^2$ for velocity).

Using  the well known expressions for the moments of the Gaussian distribution,  $ \langle \eta   \rangle = \eta_0$, $ \langle \eta^3 \rangle= \eta_0^3 + 3 \eta_0 \sigma^2$, we obtain  for the first two spatial moments of the  `Gaussian' soliton turbulence:
\be\label{gauss_mom}
\begin{split}
\overline u = -4 \kappa \langle \eta \rangle  = -4\kappa \eta_0, \\ 
\quad \overline {u^2} = \frac{16}{3} \kappa \langle \eta^3 \rangle = \frac{16}{3} \kappa (\eta_0^3 + 3 \eta_0 \sigma^2).
\end{split}
\ee
Then the turbulent wave filed variance is (cf. (\ref{var1}))
\be
\mathcal{A}^2 = 16 \kappa \eta_0^2 \left(\frac{\eta_0}{3} + \frac{\sigma^2}{\eta_0} - \kappa \right).
\ee
From the condition $\mathcal{A}^2=0$ we obtain the expression for the critial density of the Gaussian soliton gas with the mean spectral component $\eta_0$ and the spectral variance $\sigma^2$:
\be \label{var2}
\kappa_{cr}=\frac{\eta_0}{3} + \frac{\sigma^2}{\eta_0} \, .
\ee
(We recall that it was assumed that $\eta_0=O(1)$ so the persence of $\eta_0$ in the denominator is not an issue). Thus, the critical density of the `warm' gas with $\sigma>0$ is higher than  that of the  `monochromatic', cold gas, $\sigma=0$. This result has a simple physical interpretation. Consider the two-soliton interaction, which represents the basic mechanism determining macroscopic properties of the KdV soliton turbulence \cite{ek05}, \cite{pelin13}. Depending on the amplitude ratio of the interacting solitons there are three basic geometrical  configurations characterised by distinct sizes and shapes at the moment of peak interaction \cite{lax68}. The closer to each other the amplitudes of the interacting solitons are the greater  the minimum distance between their centres at the peak interaction is. This immediately leads one to the qualitative conclusion that the soliton gas consisting of solitons having a significant spectral spread  around some dominant value $\eta_0$ can acquire  greater integral density than a  gas with a narrow spectral distribution around the same value of   $\eta=\eta_0$. We note in conclusion that the choice of the Gaussian distribution  for the spectral measure $f(\eta)$ was motivated  by  the fact that it provides a transparent illustration of the difference between critical densities of the cold and warm soliton gases. The inherent restriction $\eta \ge 0$ would probably make  other distributions defined only for positive values of $\eta$ (e.g. Rayleigh or log-normal)  more relevant in the considerations of concrete physical problems. 

\section{Speed of sound in a soliton gas}
We now consder an inhomogeneous Gaussian soliton gas  by assuming in (\ref{gauss})  that $f_0= \kappa(x,t)$ but $\eta_0$ and $\sigma$ remain constant to comply with the isospectrality of the KdV evolution. 
The kinetic properties  of such a gas are fully determined by the dynamics of $\kappa(x,t)$.
Averaging the KdV conservation law $u_t+(-3u^2+ u_{xx})_x=0$ according to (\ref{kr}) we obtain
\be \label{adv_aver}
(\overline u)_t - (3 \overline{u^2})_x=0,
\ee
which, on substituting (\ref{gauss_mom}), yields the  transport equation for  the density $\kappa(x,t)$
\be \label{tr}
\kappa_t + (4 \eta_0^2 + 12 \sigma^2)\kappa_x =0.
\ee
Since $4 \eta_0^2$ is the mean velocity of the Gaussian soliton gas as a whole, the quantity $c=12 \sigma^2$  gets a natural interpretation as the `speed of sound' in a soliton gas with Gaussian spectral distribution. As expected, the `sound' does not propagate in the `cold' soliton gas with $ \sigma \to 0$ and $f \to f_0\delta(\eta - \eta_0)$. 

Remarkably, equation (\ref{tr}) is linear, so the  speed of sound in the  soliton gas does not depend on its density, which implies   the principal absence of the macroscopic wave breaking effects.  This agrees with the linearly degenerate structure of the hydrodynamic reductions of the kinetic equation (\ref{kin1}) studied in \cite{ek05}, \cite{el_jns}. Now, using the general solution of (\ref{tr}) we obtain the solution of the kinetic equation (\ref{kin1}) in the form
\be
f(\eta, x, t) = \frac{\kappa_0(\tilde x-12 \sigma^2 t)}{\sqrt{2 \pi \sigma^2}} \exp\{-\frac{(\eta - \eta_0)^2}{2\sigma^2}\},
\ee
where $\kappa_0(x)$ is the initial density distribution in the soliton gas and $\tilde x = x - 4 \eta_0^2 t$ is the transport coordinate corresponding to  the mean spectral component $\eta_0$ .

\section{Conclusions and perspectives}
We have shown that the density of KdV soliton gas is bounded from above by the value found from the condition of the vanishing for the variance (\ref{var}) for the associated random nonlinear wave field (integrable soliton turbulence).  This introduces the quantitative criterion for the notion of a dense soliton gas. The existence of the critical density  gives rise to several interesting possibilities.
One of them is related to the possible phase transitions involving soliton gas generation.   The phase transition phenomena involving soliton gases are currenly under active investigation in the context of some non-integrable dispersive systems \cite{solnyshkov, kevrekides}. In the framework of integrable systems an example of the phase transition from a smooth flow to the rapidly oscillating nonlinear regime consisting of coherent interacting solitons is well known as the DSW generation near the gradient catastrophe point but the `integrable turbulent' counterpart of this phenomenon  has not been identified yet. The second direction is related to the analysis of statistical properties of integrable soliton turbulence (PDF, power spectrum). For a rarefied gas of KdV solitons there are some recent analytical and numerical results related to the computation of skewness and kurtosis  \cite{dutykh}. The opposite limit of a dense gas, when the density is close to the critical value, could also prove analytically tractable. This is particularly compelling  in the context of the determination of the Fourier spectra of shallow water soliton turbulence observed in Ref. \cite{osborne2014} as the underlying soliton gas is dense.

 \end{document}